\documentclass[runningheads]{llncs}
\usepackage[T1]{fontenc}
\usepackage{xcolor}
\usepackage{graphicx}
\begin{document}
\title{Awayvirus: A Playful and Tangible Approach to Improve Children's Hygiene Habits in Family Education}
\titlerunning{Awayvirus: Improving Children's Hygiene Habits}
%
\author{Xiang Qi\inst{1}
\and
Yaxiong Lei\inst{2}\thanks{Corresponding Authors.}
\and
Shijing He\inst{3*}
\and
Shuxin Cheng\inst{4}
}

%
\institute{Hong Kong Polytechnic University, HKSAR, CN \email{xiang.qi@connect.polyu.hk} \and
University of St Andrews, St Andrews, UK
\email{yl212@st-andrews.ac.uk}\\
\and
King's College London, London, UK\\
\email{shijing.he@kcl.ac.uk}
\and
Central Academy of Fine Arts, Beijing, CN\\
\email{shuxincheng@cafa.edu.cn}
}
\maketitle              
\begin{abstract}
Despite various playful and educational tools have been developed to support children's learning abilities, limited work focuses on tangible toys designed to improve and maintain children's hygiene perception, habits and awareness, as well as fostering their collaboration and social abilities in home education contexts. We developed \textbf{Awayvirus} to address this research and design gap, aiming to help children gain hygiene habits knowledge through tangible blocks. Our findings indicate that a playful tangible interaction method can effectively increase children's interest in learning and encourage parents to become actively involved in their children's hygiene and health education. Additionally, Awayvirus seeks to build a collaborative bridge between children and parents, promoting communication strategies while mitigating the adverse effects of the challenging the post-pandemic period.

\end{abstract}
%
%
%


\section{Introduction}
The COVID-19 pandemic has highlighted the significance of overall health and well-being, emphasising the necessity of effective hygiene practices within family units, such as handwashing. This needs for additional measures is particularly crucial during resettlement periods, given the heightened risk of viral infections. Moreover, the pandemic has disrupted conventional education and socialisation methods for children, resulting in elevated levels of stress, and anxiety, reduced interest in learning~\cite{patel2022health}, and ineffective communication strategies~\cite{cherry-2023}. Consequently, health literacy assumes a substantial role in shaping health behaviours, especially among preschool children.

Playfulness has been recognised as an effective approach for children's education~\cite{ginsburg2007importance}. However, parents faced challenges in teaching their children proper hygiene habits using engaging and playful strategies. To address such challenges, we aim to investigate the potential of gamified educational tools with tangible interactions to enhance children's interest in learning about hygiene protection and to promote parent-child interaction during the learning process. To guide our research, we have formulated the following research questions:
\textbf{RQ1:} Can the incorporation of gamified elements increase children's interest in learning about hygiene and improve the acceptance of the educational design?
\textbf{RQ2:} How can a gamified design effectively facilitate children's learning of hand washing and good hygiene habits?
\textbf{RQ3:} In what ways can a gamified design enhance communication about hygiene between parents and children?

Our findings indicate that children's interest in learning hygiene knowledge increased after engaging with Awayvirus. We also observed that parental encouragement positively impacted children's emotions during play. Children demonstrated enthusiasm and voluntarily assisted their parents in completing collaborative educational tasks. Additionally, the power imbalance between child and parent tended to shift towards a more balanced relationship during interaction and negotiation processes.

Our main contributions are: 1) exploring how gamified educational tools with tangible interactions can increase children's learning about hygiene protection knowledge and promote parent-child communication; and 2) highlighting the challenges and findings surrounding tangible interaction in learning about hygiene protection knowledge and parent-child interaction.
\section{Related Work}
\subsection{Gamification in health professions education}
Human-computer interaction (HCI) has shown that gamification holds significant potential within health and educational research spheres~\cite{van2021gamification,sardi2017systematic}. This potential is largely due to its capacity to enhance learning outcomes and modify children's behaviour~\cite{deterding2015lens,deterding2011gamification}. Examples of such applications can be seen in efforts to alleviate stress in paediatric patients within healthcare settings~\cite{vonach2016design,liszio2020pengunaut}, as well as to bolster digital health literacy and health awareness among children~\cite{wang2022development}. Gamification has even been used to encourage physical activity and social interaction~\cite{montero2019digipack}. However, its use in fostering hygiene habits and promoting hygiene education among children remains under-explored, presenting an opportunity for further research in this domain.

\subsection{Educational Tangible Interaction}
A growing body of research highlights the extensive applicability of tangible user interfaces (TUI) in diverse environments and fields~\cite{sadka2018tangible,warren2019exploring}. The transformative potential of TUI lies in its capacity to revolutionise traditional teaching methods, enhance student engagement, and facilitate informal learning experiences~\cite{starvcivc2015transforming,marshall2007tangible,horn2018tangible}. Furthermore, the incorporation of tactile interactions in TUI has been found to support learning~\cite{devi2018augmenting}. The integration of gamified elements within TUI represents an engaging design approach, enabling children to develop essential skills through play~\cite{faber2012marbowl,alexander2019educause}.

\subsection{Parent–child interactions in home learning activities}
The crucial role of parent-child interactions in children's growth and development is well-established, particularly when it comes to enhancing learning interest through positive reinforcement~\cite{leseman1998home}. Recently, researchers have started to investigate the influence of various media, such as tablets~\cite{cingel2017parents} and artificial intelligence~\cite{druga2022family}, on these interactions. They serve as vehicles for conveying family values and beliefs to children, shaping their perceptions and interpretations of the world around them~\cite{sheehan2019parent,hiniker2018let}.

Despite these insights, research on tangible interactions specifically designed to educate early childhood learners about hygiene, with a special emphasis on parent-child interaction, remains scarce. Our design aims to bridge this gap, providing parents with a tool that captivates children's attention during home-based learning, while simultaneously stimulating their interest and fostering their communication abilities.
\section{Methodology}
Our research aims to comprehend children's perspective of hygiene habits and investigate how parental support can enhance their learning in this area. To achieve this, we employed a design thinking process~\cite{razzouk2012design} to underscore the significance of understanding adoption, usage, preference, and family health education strategies. Our multifaceted research methodology included surveys, semi-structured interviews with parents and professionals, and field trials~\cite{brown2011into}, designed to gauge collaboration, children's comprehension, and behaviours. Our study has received the ethics approval from the university's ethics committee.

\subsection{Survey}
We conducted two surveys on wjx.com, we applied social media and snowball sampling for participant recruitment. The first survey delved into the nature of parent-child interactions during game-playing, toy preferences, and educational necessities. The second survey aimed to understand parents' concerns about family education during the pandemic. We obtained 36 responses for the hygiene education strategy survey and 43 responses for the purchasing preferences survey. The results indicated a prevalent pattern among parents regarding hygiene knowledge learning and application. 24.9\% of parents expressed apprehension about their child's health protection and individual personality development within the context of home education. A majority of parents (66\%) employed varied educational tools such as animated videos (63.9\%), picture books (55.6\%), mobile games (27.8\%), and toys (47.2\%) to educate their children about hygiene and raise awareness. Yet, a significant challenge remained in effectively engaging their children in comprehending hygiene knowledge. These survey findings offered critical insights for designing educational tools that facilitate more effective children's learning about hygiene habits and knowledge.

\subsection{Interview}
We organised semi-structured interviews with two families, a 35-year-old father and his 5-year-old son, and a 30-year-old mother and her 6 years old son. We also conducted a semi-structured interview with a paediatrician from a public hospital. We inquired about their experiences, focusing on the challenges encountered in promoting family hygiene education. A shared difficulty was maintaining a balance of parental involvement across varied contexts. One father highlighted the struggle of effective communication and education during home quarantine, pointing to challenges in hygiene learning and managing distractions. Despite her nursing expertise, one mother confessed to resorting to corporal punishment due to a lack of effective educational approaches. She recognised the potential benefits of game-playing and role-playing but was concerned about excessive screen exposure. The paediatrician underscored the need to captivate preschool children's interest and cater to their inclination for hands-on learning, advocating entertaining learning methods like assembly activities, and emphasising the essential role of parental support and effective communication strategies.

The interviews revealed common issues faced by parents in promoting home hygiene education, such as limited parental experience, ineffective pedagogical approaches, power imbalance, limited child interest, and distractions. The increased shared time during home quarantine underscored the need for effective communication strategies. To overcome these challenges, the proposed design focuses on incorporating tangible and gamified design elements to increase children's engagement with hygiene knowledge, support family education, encourage parent-child communication, and enhance hygiene learning outcomes.

\subsection{Design Process}
We brainstormed and developed design concepts based on insights gleaned from previous investigations. We used a solution chart with four axes: usability, playfulness, metaphor comprehension, and prototyping practices. Subsequently, we amalgamated the four concepts into three novel product solutions and assessed each for their strengths and weaknesses. From a human-centred design perspective, we proposed tangible blocks of various geometric shapes that provide different assembly challenges and are made of natural materials~\cite{yazgin2021toys}. Our design aims to enhance children's understanding and promote their auditory and spatial language abilities~\cite{verdine2019effects}, as well as foster practical cooperation skills.

\textit{Gamification elements.}
We integrated gamification elements based on the recommendations of Hamari~\cite{hamari2014does} and Lessel~\cite{lessel2018users}. We incorporated principles including "Establishing objectives", "Capacity to overcome challenges", "Offering feedback on achievements", "Encouragement \& motivation", "Assessing \& tracking progress", "Fostering social connections" and "Fun \& playfulness". However, to prioritise parent-child interaction and learning effectiveness, we excluded competitive mechanisms and point allocation. The design solution focuses on utilising gamification elements to encourage parents and children to engage in activities such as assembling and discussing hygiene knowledge, enhancing interest and effectiveness in learning within the family context.

\begin{figure}[htbp]
\begin{center}
\includegraphics[width=0.9\textwidth]{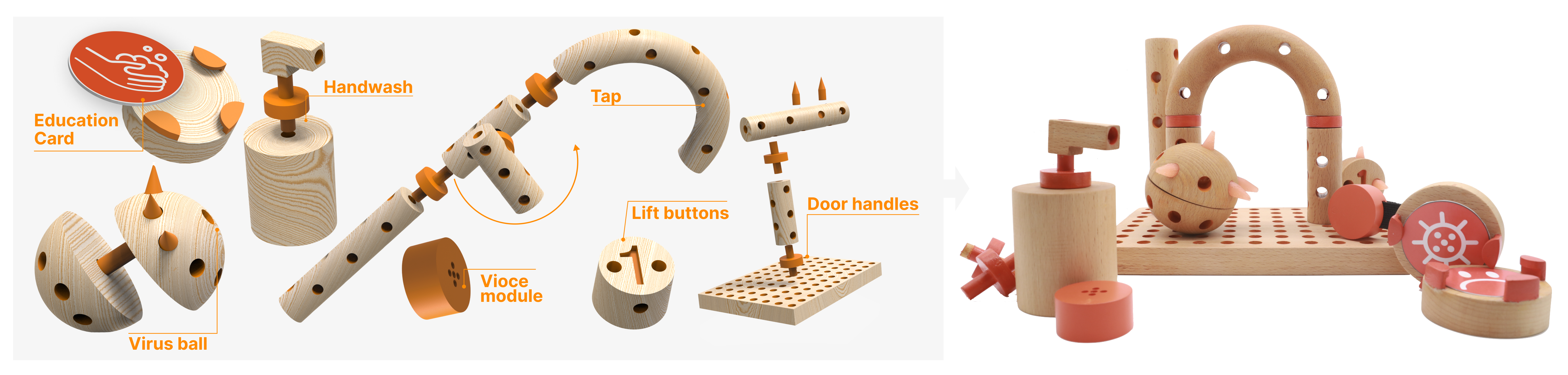}
\caption{Prototype of Awayvirus, made of different shapes and wooden materials, including tap, doorknob, handrail, and hand sanitiser. Stingers on virus blocks can be assembled onto other blocks}
\label{fig:prototype}
\end{center}
\end{figure}

\textit{Prototype.} 
Our prototype, as depicted in Figure~\ref{fig:prototype}, consists of an assembly block prototype, an Arduino motion sensor module, and learning cards. The blocks represent objects susceptible to virus attachment, and assembling them conveys the metaphor of virus transmission, such as the assembly of virus stingers onto a doorknob block. An interactive voice module with a wearable detection device prompts hand-washing by notifying the child of unhygienic behaviour. The primary objectives of Awayvirus are to raise awareness of virus attachment, emphasise timely handwashing, and facilitate learning and practice of proper handwashing techniques with parental involvement. Learning cards supplement children's understanding of basic hygiene knowledge.

\textit{Pilot Study and Results.} 
We conducted a pilot test involving a mother and her 6-year-old son in a home context, which lasted approximately 2 hours. We obtained recorded consent from the participants. The overall feedback from the tests indicated positive attitudes towards usability. However, challenges were identified, including low interest in voice interaction reminders and limited acceptance of wearable devices. For example, children displayed reluctance to wear the devices and demonstrated decreased enthusiasm. Parents also expressed concerns about small parts posing a potential choking hazard. Based on the pilot study results, we removed the voice model and wearable detection features, while the wooden building blocks and card learning components were retained.
\section{Findings}
Our findings are categorised under three main headings: Playful Learning, Learning Hygiene Habits, and Parent-Child Communication.

\begin{figure}[htbp]
\begin{center}
\includegraphics[width=9.5cm]
{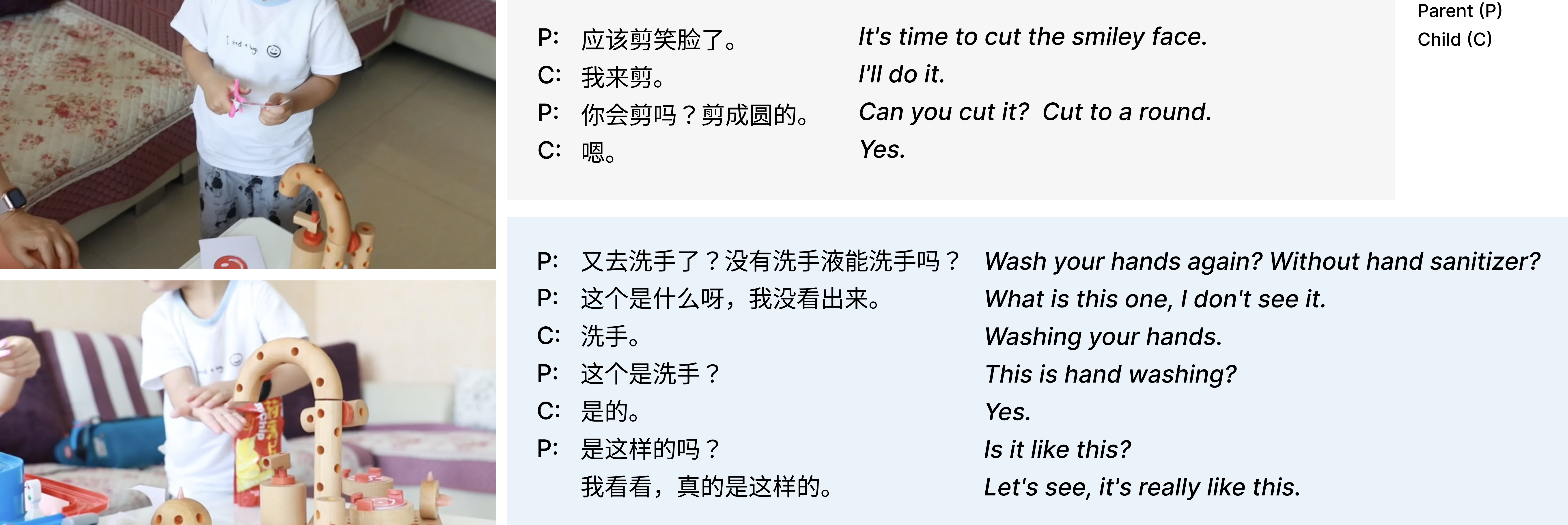}
\caption{Pilot Test in Home Context (a): Engagement for paper cutting and cooperative interaction}
\label{fig:test2}
\end{center}
\end{figure}

\textit{Playful Learning.} Tests yielded positive responses, with children showing strong enthusiasm towards the learning blocks.  We observed that children demonstrated their autonomous engagement for active participation during the cooperative learning process, such as paper cutting and simulating handwashing (see Figure~\ref{fig:test2}). The gamification elements successfully instigated intrinsic motivation and creativity, as exemplified by children modifying the learning materials based on their own ideas and activities involving adults in the handwashing process.

\begin{figure}[htbp]
\begin{center}
\includegraphics[width=9.5cm]
{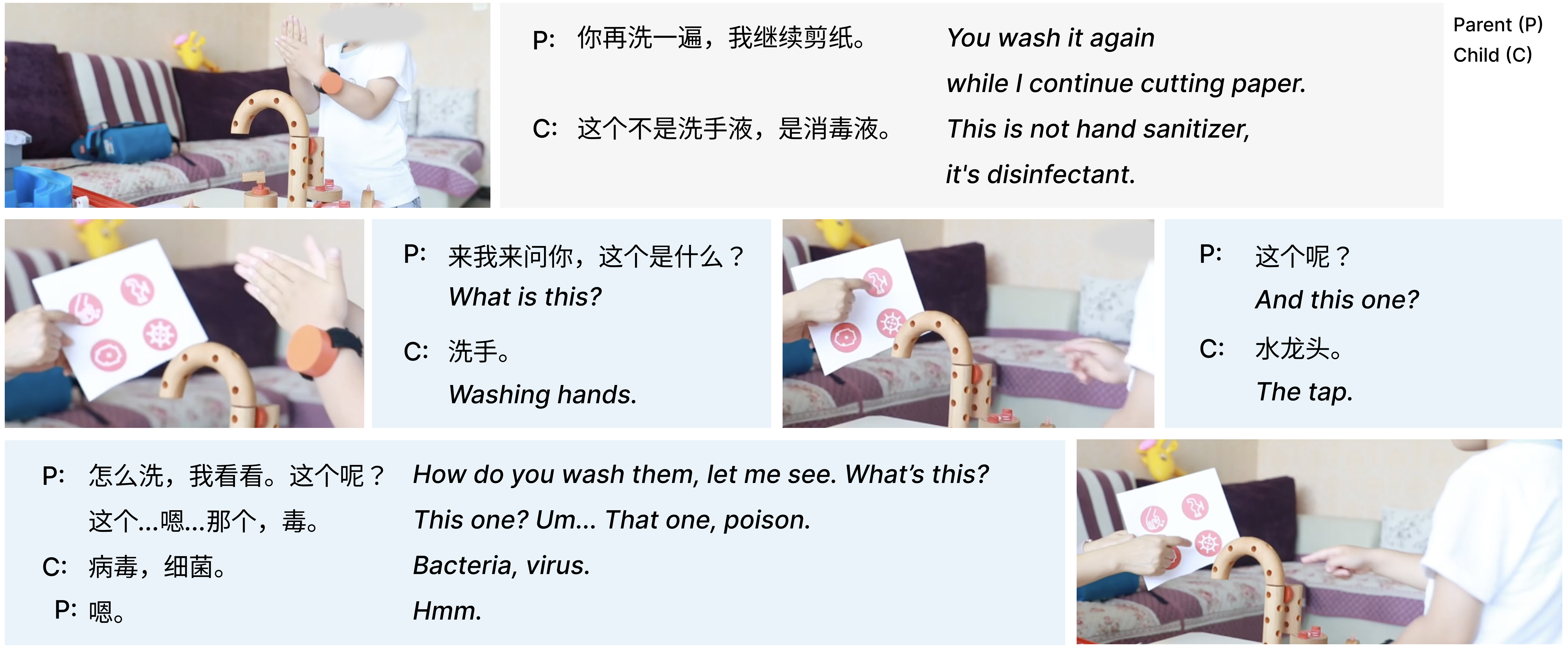}
\caption{Pilot Test in Home Context (b): The upper part shows children rectifying their parents' mistakes, and the lower part illustrates learning outcomes after playing}
\label{fig:test1}
\end{center}
\end{figure}

\textit{Learning Hygiene Habits.} Children showed a keen interest in listening and learning basic hygiene concepts through the educational cards. They also presented a propensity to imitate their parents' handwashing movements, which enhanced their understanding of proper technique. Notably, children showcased the ability to identify and articulate errors in their parents' handwashing behaviour, exemplifying their grasp of hygiene principles. While occasional difficulties in recalling specific vocabulary were observed, as shown in Figure~\ref{fig:test1}, children exhibited a solid comprehension of the association between the building blocks and real-life hygiene items, showcasing their autonomy and critical thinking skills.

\textit{Parent-Child Communication.} Incorporating playful design elements positively influenced parent-child communication and engagement. Parents observed increased eagerness and reduced distractions in their children during educational activities. The interaction during paper cutting and block assembly tasks positioned parents as guides, motivating their children (Figure~\ref{fig:test2}) and fostering the parent-child bond. Children actively participated in collaborative learning, suggesting ideas and rectifying their parents' mistakes (Figure~\ref{fig:test1}). These actions contribute to equalising power dynamics in home education. Additionally, children independently initiated further activities such as simulating handwashing and reminding parents to use hand sanitiser.

Our findings also pinpoint areas for potential improvement in the prototype design, particularly concerning safety and materials. For instance, the lack of ventilation holes in the extended parts of the virus spikes raised concerns about accidental swallowing. Adult participants stressed the importance of long-term safety considerations in selecting materials. Moreover, children exhibited a greater preference for parental interaction compared to the interactive sound and showed limited interest in the wearable device.
\section{Discussion}
In the home context, the gamified learning method stimulates children's learning interests, proactive learning attitudes, and imitating behaviours. Parents further reported that their children has been increased focus and engagement in learning about hygienic habits (e.g., handwashing), they actively involve their parents and offer feedback. However, equipping children with wearable devices led to negative emotional outbursts, which could be alleviated through parental encouragement and support. We emphasise the importance of addressing children's emotional fluctuations during learning and underscore the significant impact of parent-child relationships on children's emotional development.

We further demonstrate the efficacy of employing a playful learning approach to teach children about hygiene habits and facilitate positive parent-child communication. Future research will focus on addressing identified areas of improvement in the prototype design and refining teaching aids to enhance children's understanding and engagement. We also highlight the potential of gamification and interactive learning tools in effectively teaching essential hygiene habits to young children, fostering autonomy, and critical thinking skills, and strengthening parent-child relationships.

Our research findings highlight the role of parents in fostering collaboration with their children during joint learning activities. Awayvirus effectively maintains children's engagement and encourages them to express their ideas, seek cooperation, and address parents' improper hygiene behaviours or raise questions. The gamified design of Awayvirus promotes mutual assistance between parents and children, creates opportunities for open and equal dialogue, and mitigates parental blame and power during the learning and negotiation process. It establishes new channels for parent-child communication and enhances the overall family learning experience.

\subsection{Design Implications}
We highlight two key design insights: 1) addressing children's hands-on practice needs, and 2) fostering parent-child communication with parents as scaffolding. We also show that tangible interaction is an effective approach for early childhood educational tools, promoting interaction and collaboration.

\textit{Hands-on requirements.} Tangible interactions facilitate practical knowledge application and self-expression, aiding young children in understanding content and minimising distractions. It is essential to integrate features that accommodate children's hands-on learning needs and ensure ease of use.

\textit{Communication requirements.} Playful designs should emphasise parental involvement and cooperative processes, encouraging children's active participation and expression. The design needs to enhance parent-child communication and contributes to stronger emotional bonds and improved learning experiences.

\subsection{Limitations}
We acknowledge certain limitations in the present study. Firstly, the study was conducted during the quarantine period in China and involved a limited number of participants from a specific demographic group. Therefore, the results may not fully represent the diverse experiences and perspectives of families across different cultural, social, and economic backgrounds. Secondly, we did not investigate the long-term impact of the intervention on children's hygiene habits, behaviour, or sustainability of the observed positive outcomes. Thirdly, we did not compare the effectiveness of our prototype with other educational interventions for teaching children about hygiene protection. Future research comparing the design with other approaches could provide valuable insights into the most effective strategies for promoting hygiene habits education in early childhood.
\section{Conclusion}
Our study introduces a playful educational block approach for teaching hygiene habits to children, with the goal of stimulating their interest in learning, providing age-appropriate hygiene education, and facilitating communication between children and parents in a home-based educational context. We demonstrate the potential of the playful approach in teaching young children about hygiene habits, balancing the power differential in home-based education contexts, and promoting positive parent-child communication. Furthermore, we emphasise the importance of designing educational tools that consider children's developmental needs, engage parents, and address the emotional facets of the learning process. Moving forward, we will continue to refine and enhance our design to create an engaging, effective, and emotionally supportive learning environment for children and their families. We hope that such interventions will not only promote better hygiene habits among young children but also contribute to their overall learning development and the strengthening of familial bonds.


%
%


%
%
%
\bibliographystyle{splncs04}
\bibliography{reference}

\end{document}